\documentclass[sigconf]{acmart}

\copyrightyear{2026}
\acmYear{2026}
\setcopyright{cc}
\setcctype{by-nc-nd}
\acmConference[WWW '26]{Proceedings of the ACM Web Conference 2026}{April 13--17, 2026}{Dubai, United Arab Emirates}
\acmBooktitle{Proceedings of the ACM Web Conference 2026 (WWW '26), April 13--17, 2026, Dubai, United Arab Emirates}
\acmPrice{}
\acmDOI{10.1145/3774904.3792845}
\acmISBN{979-8-4007-2307-0/2026/04}

\usepackage{enumitem}
\usepackage{booktabs}
\usepackage{graphicx}
\usepackage{array}
\usepackage{lineno}
\usepackage{multirow}
\usepackage{hhline}
\usepackage{makecell}
\usepackage{subfigure}
\usepackage[marginal]{footmisc}
\usepackage[font=small,labelfont=bf]{caption}
\usepackage{siunitx}  
\usepackage{tcolorbox} 
\tcbuselibrary{breakable} 
\usepackage{xcolor} 
\definecolor{mygray}{HTML}{F5F9FC} 
\newtcolorbox{promptbox}{
    colback=mygray, 
    colframe=black!50, 
    arc=3pt,    
    boxrule=1pt, 
    left=6pt, 
    right=6pt,
    top=4pt,
    bottom=4pt,
    boxsep=2pt,
    fontupper=\small, 
    before skip=8pt, 
    after skip=8pt, 
    width=\linewidth, 
    breakable 
}

\AtBeginDocument{%
  \providecommand\BibTeX{{%
    \normalfont B\kern-0.5em{\scshape i\kern-0.25em b}\kern-0.8em\TeX}}}

\copyrightyear{2025}
\acmYear{2025}

\begin{document}

\title{
DOS: Dual-Flow Orthogonal Semantic IDs for Recommendation in Meituan
}


\author{Junwei Yin}
\authornote{Both authors contributed equally to this research.}

\affiliation{%
  \institution{Meituan}
  \city{Chengdu}
  \country{China}
}
\email{yinjunwei03@meituan.com}

\author{Senjie Kou}
\authornotemark[1]

\affiliation{%
  \institution{Meituan}
  \city{Chengdu}
  \country{China}\textbf{}
}
\email{kousenjie@meituan.com}

\author{Changhao Li}
\affiliation{%
\institution{Meituan}
   \city{Chengdu}
  \country{China}
  }
\email{lichanghao@meituan.com}

\author{Shuli Wang}
\authornote{Corresponding author.}
\affiliation{%
  \institution{Meituan}
   \city{Chengdu}
  \country{China}
}
\email{wangshuli03@meituan.com}

\author{Xue Wei}
\affiliation{%
\institution{Meituan}
   \city{Chengdu}
  \country{China}
  }
\email{weixue06@meituan.com}

\author{Yinqiu Huang}
\affiliation{%
\institution{Meituan}
   \city{Chengdu}
  \country{China}
  }
\email{huangyinqiu@meituan.com}

\author{Yinhua Zhu}
\affiliation{%
\institution{Meituan}
   \city{Chengdu}
  \country{China}
  }
\email{zhuyinhua@meituan.com}

\author{Haitao Wang}
\affiliation{%
\institution{Meituan}
   \city{Chengdu}
  \country{China}
  }
\email{wanghaitao13@meituan.com}

\author{Xingxing Wang}
\affiliation{%
\institution{Meituan}
   \city{Beijing}
  \country{China}
  }
\email{wangxingxing04@meituan.com}

\renewcommand{\shortauthors}{Junwei Yin et al.}



\begin{abstract}
Semantic IDs serve as a key component in generative recommendation systems. They not only incorporate open-world knowledge from large language models (LLMs) but also compress the semantic space to reduce generation difficulty. However, existing methods suffer from two major limitations: (1) the lack of contextual awareness in generation tasks leads to a gap between the Semantic ID codebook space and the generation space, resulting in suboptimal recommendations; and (2) suboptimal quantization methods exacerbate semantic loss in LLMs. To address these issues, we propose Dual-Flow Orthogonal Semantic IDs (DOS) method. Specifically, DOS employs a user–item dual-flow framework that leverages collaborative signals to align the Semantic ID codebook space with the generation space. Furthermore, we introduce an orthogonal residual quantization scheme that rotates the semantic space to an appropriate orientation, thereby maximizing semantic preservation. Extensive offline experiments and online A/B testing demonstrate the effectiveness of DOS. The proposed method has been successfully deployed in Meituan’s mobile application, serving hundreds of millions of users.

\end{abstract}





\maketitle

\section{Introduction}
Personalized recommendation in industrial settings involves selecting items from a candidate set of hundreds of millions. Traditional systems based on large-scale item-ID vocabularies pose a fundamental challenge to recommendation efficiency and parameter scaling. In contrast, the emerging paradigm of Generative Recommendation (GR), which utilizes Semantic IDs (SID), shows considerable promise for streamlining the recommendation pipeline and advancing scaling laws\cite{actionpiece}. Semantic IDs serve as a critical module within this framework, notable for their ability to integrate the open-world knowledge from LLMs while simultaneously compressing the semantic space \cite{rpg} to alleviate the complexity of generation.

Despite their promise, current approaches to learning Semantic IDs remain suboptimal, primarily due to two inherent limitations.
\textbf{First, the Codebook-Generation Gap}: Most existing methods learn Semantic IDs in a task-agnostic manner\cite{tiger,rpg}, often relying solely on reconstruction objectives. This approach overlooks the critical contextual information of the final generation task. Consequently, a significant gap emerges between the obtained Semantic ID codebook space and the generation space of the downstream task. This misalignment forces the generative model to operate with a suboptimal representation, ultimately leading to degraded recommendation quality.
\textbf{Second, Semantic Loss in Quantization}: The quantization of continuous semantic representations into discrete tokens inevitably incurs information loss. While effective in other domains, prevailing quantization methods are not tailored to preserve the nuanced semantic structures essential for LLM comprehension, as they are primarily designed for \textbf{clustering} items \cite{saviorrec} or user-item interactions  \cite{DAS}. This unsuitable quantization method exacerbates semantic distortion, hindering the LLM's ability to accurately interpret the intended meaning of the Semantic IDs.

\begin{figure*}[htbp]
    \centering
    \includegraphics[width=0.8\linewidth,height=0.3\textwidth]{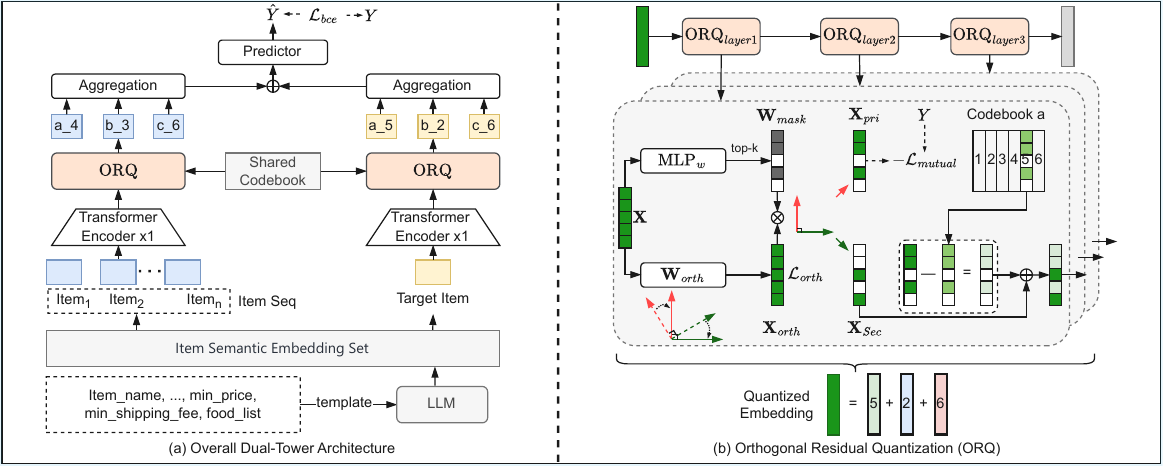}
    \caption{Overview of DOS. The left part (a) depicts the overall dual-flow architecture for processing the user's click sequence and the target item. The right part (b) illustrates the ORQ module, which leverages $\mathbf{X}_{orth}$ and $\mathbf{W}_{mask}$ to disentangle primary ($\mathbf{X}_{pri}$) and secondary ($\mathbf{X}_{sec}$) semantic features through the joint optimization of $\mathcal{L}_{Orth}$ and $\mathcal{L}_{Mutual}$, enabling progressive selection of task-relevant semantics.
    }
    \label{fig1}
\end{figure*}

To overcome these challenges, we propose a novel framework named Dual-Flow Orthogonal Semantic IDs (DOS). Our approach is designed to explicitly align the learning of SIDs with the objective of generative recommendation.
Specifically, DOS introduces a user-item dual-tower architecture that leverages collaborative signals from an explicitly and implicit perspective. This design ensures that the learned SID codebook space is inherently aligned with the generative space of the LLM, effectively bridging the gap left by task-agnostic methods.
This novel quantization method strategically rotates the semantic space to an orientation that minimizes information loss during discretization, thereby maximally preserving the semantic fidelity of the original representations.
The main contributions of our DOS are summarized as follows:
\begin{itemize}[leftmargin=*, topsep=0pt, partopsep=0pt]
\item We propose a Dual-Flow Integration (DFI) architecture that explicitly models user-item interactions and implicitly aligns their semantic spaces via a shared codebook.
\item We devise an Orthogonal Residual Quantization (ORQ) module that iteratively extracts task-relevant semantics across layers to minimize information loss.
\item Extensive offline experiments and online A/B tests on large-scale datasets from Meituan's platform demonstrate that DOS significantly outperforms state-of-the-art baselines.
\end{itemize}

\section{Approach}
To bridge the gap between SID codebook space and generation space, we propose DOS, a novel model that performs orthogonal residual quantization within a dual-flow architecture (Figure \ref{fig1}). This architecture integrates user and item information through a two-fold process: modeling their direct interactions and aligning their representations via a shared codebook into a unified semantic space. Within this process, the ORQ module progressively quantizes the most task-relevant semantics for effective downstream prediction.

\subsection{Dual-Flow Integration (DFI)}
In this process, DOS leverages a dual-flow architecture to integrate collaborative information from the user's click sequence and target item directly into the quantization operation.

\textbf{Semantic Embedding Generation}. We generate comprehensive semantic item representations by designing a structured prompt template that directs an LLM (Qwen3-0.6B-embedding) to produce embeddings as follows. 
\begin{promptbox}
\texttt{\{Item name\}} is a \texttt{\{second category\}} restaurant specializing in \texttt{\{third category\}}. Reservations are not accepted / are accepted. The minimum order amount is \texttt{\{min price\}} yuan, and the minimum delivery fee is \texttt{\{min shipping fee\}} yuan. Featured dishes include \texttt{\{dish 1, dish 2, ..., dish n\}}.
\end{promptbox}
These representations are stored as a set $\mathbf{E}_{semantic} = \{ e_1, e_2, \cdots, \\ e_N \} \in \mathbb{R}^{N \times d} $ for subsequent use, where $N$ indicates the total number of items and $d$ is the embedding dimension.

\textbf{User-Item Information Integration}. We integrate user-item collaborative information in two folds: explicitly modeling each user’s behavioral sequence and implicitly leveraging a shared codebook that encodes global collaborative patterns.

Explicitly, to alleviate the gap between item quantification and generation tasks,  we directly model the relationship between the user’s click sequence and the target item during quantization. This is achieved by first retrieving the semantic embeddings of both from the precomputed set $\mathbf{E}_{\text{semantic}}$ and then encoding them with a one-layer Transformer encoder.
\begin{align}
    \mathbf{X}_{user} = \mathrm{Encoder}_{user}(\mathbf{X}_{seq}), \hspace{0.3cm}  \mathbf{X}_{item} = \mathrm{Encoder}_{item}(\mathbf{X}_{target}), \label{eq1}
\end{align}
where $\mathbf{X}_{user} = [e_m, e_n, \cdots, e_p] \in \mathbb{R}^{S \times d}$ and $\mathbf{X}_{item} = [e_t, e_t, \ldots , e_t] \in \mathbb{R}^{S \times d}$ are the representations obtained by the encoders. Here, $m, n, p, \\t \in {1, 2, \cdots, N}$ are item indices and $S$ denotes the sequence length.
$\mathbf{X}_{user}$ and $\mathbf{X}_{item}$ are then processed by the ORQ module to extract task-relevant semantics at three hierarchical levels:
\begin{align}
    \mathcal{C}_{user} = \mathrm{ORQ}_{user}(\mathbf{X}_{user}), \hspace{0.6cm}  \mathcal{C}_{item} = \mathrm{ORQ}_{item}(\mathbf{X}_{item}), \label{eq2}
\end{align}
where $\mathcal{C}_{user}$ and $\mathcal{C}_{item}$ are quantified semantic representations.

Finally, the three-level quantized representations are aggregated into final embeddings, which are then concatenated for the downstream prediction:
\begin{align}
    \mathbf{E}_{user} =\hspace{0.1cm} \mathrm{Agg}&\mathrm{re}_{user}(\mathcal{C}_{user}), \hspace{0.4cm}  \mathbf{E}_{item} = \mathrm{Aggre}_{item}(\mathcal{C}_{item}), \label{eq3} \\
    &\hat{\mathbf{Y}} = \mathrm{MLP}_{pred}(\mathbf{E}_{user} \oplus \mathbf{E}_{item}), \label{eq4} \\
    \mathcal{L}_{BCE} = &-\sum_{i=1}^{Z} [y_i log(\hat{y}_i)+(1-y_i)log(1-\hat{y}_i)], \label{eq5}
\end{align}
where $y_i \in \mathbf{Y}$ and $\hat{y}_i \in \hat{\mathbf{Y}}$ indicate the ground-truth and the predicted labels, respectively. 

Implicitly, we leverage a shared codebook to integrate the relationship between the user's behavior sequence and the target item. This approach is effective due to the homogeneous nature of their representations: both are derived from items. By quantizing them with the same codebook, we project both the user's interest (inferred from the sequence) and the target item into a unified semantic space. Consequently, during downstream prediction, this shared coding scheme ensures that their representations enable effective comparison, facilitating accurate matching.

\subsection{Orthogonal Residual Quantization (ORQ)}
Although LLMs produce semantically rich representations, they are often generic and noisy for specific tasks. To distill task-relevant features, we propose an Orthogonal Residual Quantization (ORQ) module. ORQ operates layer-wise: at each stage, it quantizes primary features for the current granularity, while passing secondary features and residuals to the next layer. This iterative process effectively distills relevant information at the appropriate level while preserving finer-grained signals. Here we take $ORQ_{layer1}$ as an example to elaborate its realization, as illustrated in Figure \ref{fig1} (b).

\textbf{Feature Orthogonality}. The input $\mathbf{X}$ is transformed by an orthogonal matrix to an optimal orientation, preserving its full semantic content. 
\begin{align}
    \mathbf{X}_{orth} = \mathbf{W}_{orth} \mathbf{X}, \hspace{0.3cm}
    \mathcal{L}_{Orth} = \left\|\mathbf{W}_{orth} \mathbf{W}^{T}_{orth}-\mathbf{I}\right\|^{2}, \label{eq6}
\end{align}
where $\mathbf{X}_{orth}$ represents the rotated $\mathbf{X}$ and $\mathcal{L}_{orth}$ denotes the orthogonal loss used to constrain the rotation angle. $\mathbf{W}_{orth}$ indicates the orthogonal matrix and $\mathbf{I}$ is an identity matrix.

Meanwhile, to justify the dimensions that benefit for downstream task, we introduce a two layer MLP ($\mathrm{MLP}_{w}$) to generate a weight score ($\mathbf{W}_{score}$) for each dimension. Then, we select top-$k$ salient dimension as the primary feature $\mathbf{X}_{pri}$ and the left as the secondary feature $\mathbf{X}_{sec}$ as follows:

\begin{equation}
\mathbf{W}_{score} = \mathrm{Sigmoid}(\mathrm{MLP}_{w}(\mathbf{X})), \label{eq7} \hspace{0.1cm}
\mathbf{W}_{mask}(i) =
\begin{cases}
1, & \text{if } i \in \mathcal{I}_{\text{top-}k} \\
0, & \text{otherwise},
\end{cases}
\end{equation}
where $\mathcal{I}_{\text{top-}k}$ is the set of indices corresponding to the $k$ largest weight scores, and $\mathbf{W}_{mask}$ indicates the mask matrix used to select primary features:
\begin{align}
    \mathbf{X}_{pri} &=\mathbf{W}_{mask} \mathbf{X}_{orth}, \hspace{0.3cm} \mathbf{X}_{sec} = (\mathbf{1}-\mathbf{W}_{mask}) \mathbf{X}_{orth}, \label{eq8} \\
    \mathcal{L}&_{Mutual} = -\sum_{\mathbf{X}_{pri},\mathbf{Y}} \mathbb{P}(\mathbf{X}_{pri}, \mathbf{Y})log\frac{\mathbb{P}(\mathbf{X}_{pri}, \mathbf{Y})}{\mathbb{P}(\mathbf{X}_{pri}), \mathbb{P}(\mathbf{Y})}, \label{eq9}
\end{align}
where $\mathbf{X}_{pri}$ is orthogonal to $\mathbf{X}_{sec}$ since $\mathbf{X}_{pri} \cdot \mathbf{X}_{sec} = 0$, which guarantees all the primary feature be keeped. To ensure the primary features are task-relevant, we use mutual information to measure its relevance to the task label $\mathbf{Y}$.

\textbf{Residual Quantization}. We perform residual quantization on the selected primary features, and then we concatenate residual feature and secondary feature as the input of next layer ($X_{next}$):
\begin{align}
    \mathbf{X}_{resi} =\mathbf{X}_{pri} - \mathcal{C}_{i}, \hspace{0.4cm} \mathbf{X}_{next} = \mathbf{X}_{sec} \oplus \mathbf{X}_{resi}, \label{eq9}
\end{align}
where $\mathbf{X}_{resi}$ indicates the residual embedding after quantization and $\mathcal{C}_{i}$ is the $i$-th codebook vector that most similar with $\mathbf{X}_{pri}$.


Finally, we train DOS by jointly minimizing the losses as follows.
\begin{equation}
\left\{\begin{array}{l}
\mathcal{L}=\mathcal{L}_{\text {BCE}}+\alpha (\mathcal{L}_{\text{Orth}}+\mathcal{L}_{Mutual})+\mathcal{L}_{\text {Recon}}+\mathcal{L}_{\text {VQ}}, \text { where } \\
\mathcal{L}_{\text {Recon}}=\|\mathbf{E}-\mathbf{\hat{E}}\|^{2} \\
\mathcal{L}_{\text {VQ}}=\sum_{i=1}^{L}\left\|sg\left[\mathbf{X}_{resi_{l-1}}\right]-\mathcal{C}_{i_{l}}\right\|^{2}+\beta\left\|\mathbf{X}_{resi_{l-1}}-sg\left[\mathcal{C}_{i_{l}}\right]\right\|^{2},
\end{array}\right.
\end{equation}
where $\alpha$ and $\beta$ are hyperparameters used for balancing and $sg$ means stop-gradient operation.

\section{Experiments}\label{Experiments}
\subsection{Experiment Settings}
\subsubsection{Datasets and Baselines}
To ensure alignment with our online objectives, we followed prior methods \cite{DAS},\cite{pcr-ca} by constructing the evaluation dataset solely from production data, forgoing incompatible public datasets.

\textbf{Quantization Dataset}: To ensure comprehensive coverage of business items, we collected user behavior records from the past year, encompassing 24 million unique items. For model training, we constructed 50-length user click sequences. The item immediately following each sequence (the 51st) was treated as a positive sample (label=1), while negative samples (label=0) were generated through random sampling.
\textbf{Offline Training Dataset}: We used 60 days of online data, reserving the most recent 3 days for validation and the preceding 57 days for training, totaling 180 million interactions. We select rq-vae \cite{RQ-VAE}, rq-kmeans\cite{qarm}, DAS\cite{DAS}, and HSTU\cite{hstu} as baselines for comparison.

\subsubsection{Implementation Details}
The quantization dataset is split into training, validation, and test sets in an 8:1:1 ratio. The parameters are configured as follows: the hyperparameters are set to $\alpha=0.1$ and $\beta=0.25$, semantic embedding dimension $d=1024$, codebook layers $L=3$, the codebook size for each layer is 1024, and batch size is 1024. Training employs early stopping with a patience of 5. For evaluation, we use the F1 score and AUC for the quantization task, and the Hit@10 for the generation task.

\subsection{Performance Comparison}\label{exp_result}
We evaluate DOS on generative and downstream tasks. For a fair comparison, the baseline methods RQ-KMeans and RQ-VAE are fine-tuned with the application metrics, as they lack this alignment. The input to these baselines is the user's click history concatenated with the target item. 

\subsubsection{Downstream Task Comparison}
As illustrated in Table \ref{table1}, our DOS achieves the best downstream task performance. RQ-KMeans performs the worst because it relies on a static clustering mechanism, which lacks a learnable component. In contrast, RQ-VAE improves upon this by employing an encoder-decoder architecture to learn the quantization. DAS achieves inferior performance compared to our proposed DOS. This is because DAS aligns the user and item quantization processes indirectly through a collaborative information module, as opposed to the more integrated approach in DOS that directly models user behavior sequences with the target item. The relative indirectness of DAS's alignment may introduce semantic space deviation. Our DOS achieves even greater performance by utilizing an encoder coupled with an ORQ module, which proactively selects the most task-relevant semantic information. The rationale for not incorporating a decoder in DOS is elaborated upon in the ablation study (Section \ref{3.3}).

\begin{table}[h]
\centering
\setlength{\tabcolsep}{4pt} 
\footnotesize
\caption{Performance Comparison with Respect to Downstream Task Metrics (AUC and F1-Score).}
\label{table1}
\begin{tabular}{lcccc}
\hline
Model & RQ-KMeans & RQ-VAE & DAS & DOS \\
\hline
AUC & 0.8363 & 0.8526 & \underline{0.8539} & \textbf{0.8763}\\
F1-Score & 0.7641 &  0.7739 & \underline{0.7869}  & \textbf{0.8057}\\
\hline
\end{tabular}
\end{table}

\subsubsection{Downstream Next Token Prediction Performance}
The quality of the SID generated by our DOS method is ascertained through evaluation within the HSTU framework. Model performance is benchmarked against the Hit@10 metric on both the overall dataset and four distinct business types (Busi\_A to Busi\_D). The results, summarized in Table \ref{table2}, indicate that HSTU-DOS achieves significantly superior performance over HSTU-RQ-KMeans and HSTU-DAS. This finding is consistent with the downstream task results in Table \ref{table1}, providing strong evidence for the enhanced efficacy of DOS in producing high-quality SID.

\begin{table}[h]
\centering
\setlength{\tabcolsep}{4pt} 
\footnotesize
\caption{Hit@10 Performance by Business Category.}
\label{table2}
\begin{tabular}{lccccc}
\hline
Model & All & Busi\_A & Busi\_B & Busi\_C  & Busi\_D  \\
\hline
HSTU-RQ-KMeans & 0.0410 & 0.0252 & 0.0554 & 0.0398 & 0.0421 \\
HSTU-DAS & \underline{0.0511} & \underline{0.0325} & \underline{0.0672} & \underline{0.0502} & \underline{0.0541} \\
HSTU-DOS & \textbf{0.0676} & \textbf{0.0457} & \textbf{0.0797} & \textbf{0.0730} & \textbf{0.0718} \\
\hline
\end{tabular}
\end{table}

\subsection{Ablation Study} \label{3.3}
To validate the design choices of our proposed DOS framework, we conducted comprehensive ablation studies. We evaluated three variants: ``-w MLP-Encoder'' (replacing the Transformer encoder with a two-layer MLP), ``-w/o Share'' (not sharing codebook), and ``-w Decoder'' (adding a decoder for reconstruction). As detailed in Table \ref{table3}, all modifications lead to performance degradation. The MLP encoder fails to model sequence-item relationships effectively, while removing the shared codebook misaligns the semantic spaces of user behaviors and the target item. The performance drop from the added decoder stems from a conflict between its reconstruction objective and the goal of selecting task-relevant information.

\begin{table}[h]
\centering
\setlength{\tabcolsep}{4pt} 
\footnotesize
\caption{Ablation Study Results on AUC and F1-Score.}
\label{table3}
\begin{tabular}{l|cccc}
\toprule
Metric & -w MLP-Encoder & -w/o Share & -w Decoder & DOS \\
\midrule
AUC & 0.8462 & \underline{0.8671} & 0.8626 & \textbf{0.8763} \\
F1-Score & 0.7645 & \underline{0.7950} & 0.7862 & \textbf{0.8057} \\
\bottomrule
\end{tabular}
\end{table}

\subsection{Online Experiment Results}
To quantify the incremental gain from the SID generated by our DOS framework in a live environment, we conducted a one-week online A/B test using 30\% of Meituan's production traffic. The experiment was designed to compare the original baseline model against a variant where SID were directly incorporated without any other architectural modifications.

The online A/B test demonstrated the effectiveness of our DOS-generated Semantic ID (SID), which directly contributed to a 1.15\% increase in online revenue. This result provides strong validation that the SID effectively captures task-relevant semantic information, leading to tangible business improvements.

\section{Conclusion}
In this paper, we propose Dual-flow Orthogonal Semantic IDs (DOS), a network devised to enhance generative recommendation by learning context-aware and semantically-rich compact representations. DOS models user–item interactions within a dual-flow integration framework (DFI) that aligns the Semantic ID codebook space with the generative space using collaborative signals. Specifically, we introduce an orthogonal residual quantization (ORQ) module, which rotates the semantic space to an optimal orientation to minimize information loss during quantization. Moreover, the orthogonal decomposition separates primary and residual semantic components, thereby maximizing semantic preservation and improving alignment for recommendation tasks. Extensive experiments on large-scale online A/B tests within Meituan’s platform demonstrate the superiority of DOS in generative recommendation.
\bibliographystyle{ACM-Reference-Format}
\bibliography{main}

\end{document}